\documentclass[fp,twocolumn]{jpsj3}

\usepackage{bm}
\usepackage{graphicx}
\usepackage{newtxtext,newtxmath}
\usepackage{amsmath,amssymb}
\usepackage[up]{subfigure}
\usepackage{braket}

\graphicspath{{./fig/}}

\title{Performance Enhancement of Quantum Annealing under the Lechner-Hauke-Zoller Scheme by Non-linear Driving of the Constraint Term}

\author{Yuki Susa$^1$\thanks{y-susa@nec.com} and Hidetoshi Nishimori$^{2,3,4}$}

\inst{
$^1$System Platform Research Laboratories, NEC Corporation, Tsukuba, Ibaraki 305-8501, Japan \\
$^2$Institute of Innovative Research, Tokyo Institute of Technology, Yokohama, Kanagawa 226-8503, Japan \\
$^3$Graduate School of Information Sciences, Tohoku University, Sendai, Miyagi 980-8579, Japan \\
$^4$RIKEN Interdisciplinary Theoretical and Mathematical Sciences (iTHEMS), Wako, Saitama 351-0198, Japan
}

\abst{
We analyze the performance of quantum annealing as formulated by Lechner, Hauke, and Zoller (LHZ), by which a Hamiltonian with all-to-all two-body interactions is reduced to a corresponding Hamiltonian with local many-body interactions.  Mean-field analyses show that problematic first-order quantum phase transitions that exist in the original LHZ formulation can be avoided, and thus an exponential speedup is achieved, if we drive the coefficient of the many-body term, which represents the constraint, non-linearly as a function of time.  This result applies not only to a simple ferromagnetic model but also to the spin glass problem if a parameter in the spin glass model is chosen appropriately. 
}

\begin{document}
\maketitle
\section{Introduction}
Quantum Annealing (QA) is a metaheuristic for solving combinatorial optimization problems using quantum fluctuations~\cite{kadowaki1998quantum, brooke1999quantum, santoro2002theory, santoro2006optimization, das2008colloquium, morita2008mathematical,hauke2019perspectives}, and is closely related with adiabatic quantum computation~\cite{farhi2001quantum,albash2018adiabatic}. The goal of QA is to obtain the ground state of a classical Ising model, to which a combinatorial optimization problem can be reduced \cite{lucas2014ising}. It is usually the case that the resulting Ising Hamiltonian has long-range interactions including all-to-all interactions, but the current annealing device, the D-Wave quantum annealer, does not directly implement long-range interactions. One therefore has to employ the procedure of embedding \cite{Choi2008,Choi2011} to represent long-range interactions in terms of a combination of short-range interactions.   This leads to an overhead in the qubit count and also tends to cause errors induced by imperfect realization of embedding in the real device.

Lechner, Hauke, and Zoller (LHZ) \cite{lechner2015quantum} proposed an ingenious scheme to partly mitigate the above problems by mapping all-to-all interactions to single-qubit terms in the Hamiltonian supplemented by local four-body interactions introduced to guarantee the equivalence of two formulations. Although the issue of overhead in the qubit count still exists, the reduction of all-to-all interactions to a local representation is certainly advantageous in the device implementation as well as from the viewpoint of mitigation of errors caused by imperfections in embedding. Several proposals have been made to realize the LHZ scheme \cite{leib2016transmon,chancellor2017circuit,puri2017engineering,puri2017quantum,goto2019quantum}.

There have also been attempts to analyze the LHZ scheme theoretically.  Leib, Zoller, and Lechner \cite{leib2016transmon} proposed a method to reduce four-body interactions in the LHZ Hamiltonian to two-body interactions by introducing auxiliary qubits. They also showed that a proper control of the coefficient of the constraint terms is likely to improve the performance. Hartmann and Lechner \cite{hartmann2019quantum} used non-stoquastic counter-diabatic drivers for better performance, and the same authors recently introduced a mean-field like method to analyze the effect of inhomogeneity in the transverse field for increased success probabilities \cite{hartmann2019rapid}.

We have been inspired by these developments and have analyzed the LHZ scheme within the framework of mean-field theory. The result shows that a non-linearity in the coefficient of the constraint term as a function of time leads to avoidance of first-order phase transitions that exist in the original linear time dependence of the coefficient.  This implies an exponential speedup from the view point of adiabatic quantum computation because a first-order phase transition usually accompanies an exponentially-closing energy gap as a function of the system size, meaning an exponential computation time according to the adiabatic theorem of quantum mechanics \cite{jansen2007,lidar2009}. 

This paper is organized as follows. After an introduction to the LHZ model and its mean-field version in Sec. \ref{section:LHZ_model}, 
we analyze the problem analytically in Sec. \ref{section:MF_numerical}. Conclusion is described in Sec. \ref{section:conclusion}.

\section{The LHZ Model and the Mean-field Model}
\label{section:LHZ_model}

The conventional QA has the Hamiltonian
\begin{align}
\label{eq:qa_hamiltonian}
\hat{H}(s) =s \hat{H}_{P} + \left(1-s\right) \hat{V},
\end{align}
where $\hat{H}_P$ is the Ising Hamiltonian representing a combinatorial optimization problem,
\begin{align}
    \hat{H}_P=-\sum_{i\ne j} J_{ij} \hat{\sigma}_i^z\hat{\sigma}_j^z-\sum_{i} h_i \hat{\sigma}_i^z
    \label{eq:Ising_model0}
\end{align}
and $\hat{V}$ is the transverse field to induce quantum fluctuations,
\begin{align}
    \hat{V}=-\sum_i \hat{\sigma}_i^x
\end{align}
with $\hat{\sigma}_i^{z(x)}$ denoting the $z$($x$) component of the Paul operator at site (qubit) $i$. The parameter $s=t/T$ is the normalized time running from 0 to $1$ as the time $t$ proceeds from 0 to $T$, and thus $T$ is the total computation time (annealing time).

One starts at $s=0$ from the trivial ground state of $\hat{V}$ and increases $s$ with the expectation that the ground state of the Ising Hamiltonian $\hat{H}_P$ is reached at $s=1~(t=T)$.  According to the adiabatic theorem of quantum mechanics, the computation time $T$ necessary for the system to stay close enough to the instantaneous ground state is proportional to the inverse polynomial of the minimum energy gap $\Delta =\min_{s}\{E_1(s)-E_0(s)\}$, where $E_0(s)$ and $E_1(s)$ are the instantaneous ground and first-excited state energies, respectively.  If this energy gap closes exponentially as a function of the system size $N_l$ (the number of logical qubits) as is usually the case at a first-order quantum phase transition, the computation time $T$ is grows exponentially $e^{aN_l}~(a>0)$, which means that the problem is hard to solve by QA. It is therefore highly desirable to avoid or remove first-order phase transitions. 

\begin{figure}[tb]
\centering
	\includegraphics[width=0.3\textwidth]{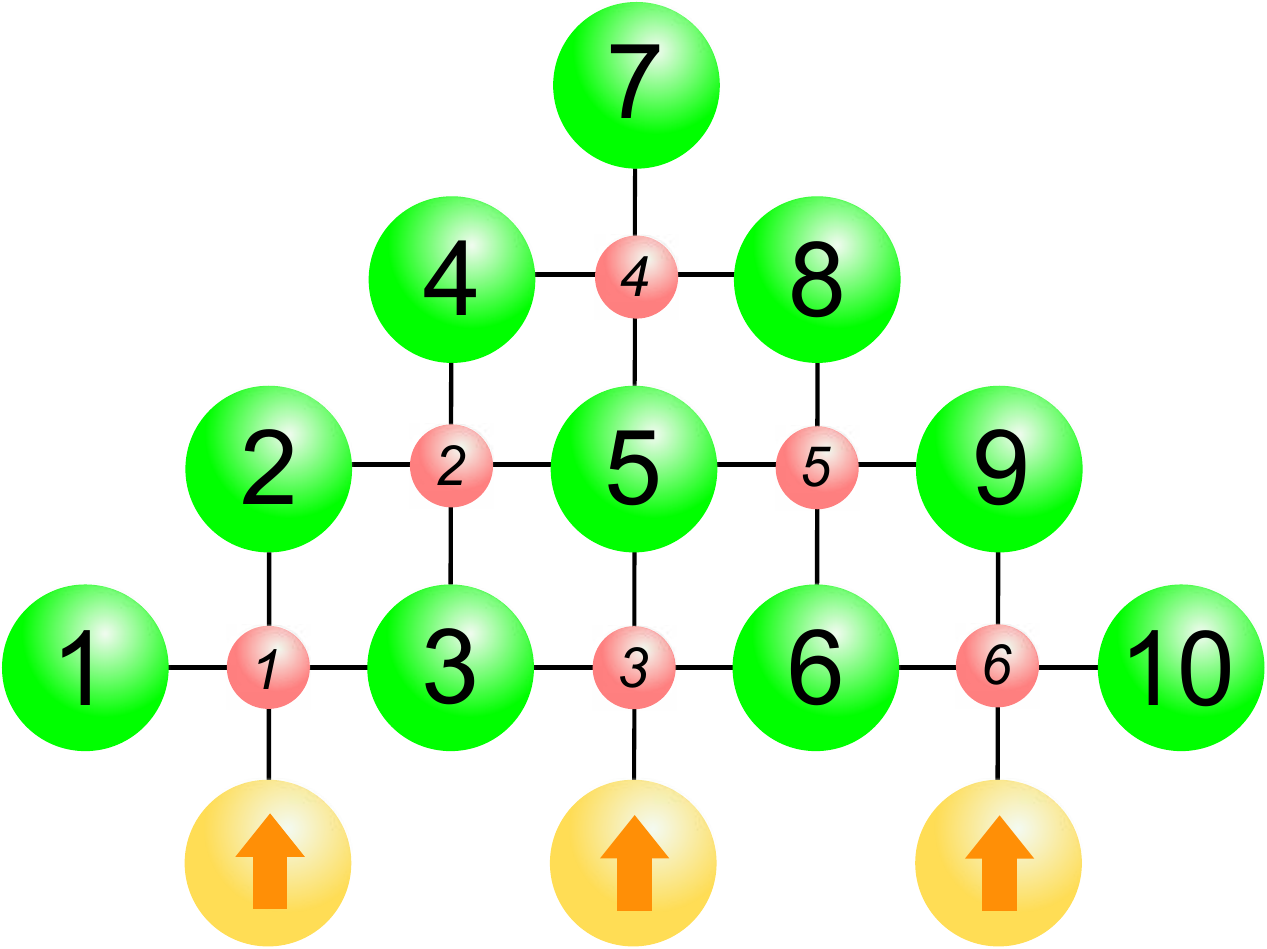}
\caption{(Color online) Qubit configurations in the LHZ model for $N_l=5$ logical qubits. Large green circles denote physical qubits and the four qubits around each small red circle (plaquette) consist a four-body interaction. The number in a green or red circle is the index of a physical qubit $k$ or a plaquette $l$, respectively. The state of auxiliary physical qubit at the bottom row (yellow) is fixed to $\ket{\uparrow}$.}
  \label{fig1}
\end{figure}

The LHZ scheme \cite{lechner2015quantum} reduces the all-to-all interactions implied in the Ising Hamiltonian eq. (\ref{eq:Ising_model0}) to single-body terms supplemented by four-body constraint terms to enforce  equivalence to the original problem,
\begin{align}
\label{eq:lhz}
\hat{H}_{P_1} &= - \sum_{k=1}^{N} J_k \hat{\sigma}_k^z -\sum_{l=1}^{N_c} \hat{\sigma}_{(l,n)}^z \hat{\sigma}_{(l,w)}^z \hat{\sigma}_{(l,s)}^z \hat{\sigma}_{(l,e)}^z
\end{align}
through the correspondence
\begin{align}
    J_{ij} \hat{\sigma}_i^z\hat{\sigma}_j^z \longrightarrow J_k\hat{\sigma}_k^z.
\end{align}
The number of physical qubits $N$ in the LHZ Hamiltonian eq. (\ref{eq:lhz}) is the number of all-to-all interactions in the original model,
\begin{align}
    N=\frac{1}{2}N_l (N_l-1),
\end{align}
and the number of constraints in the second term on the right-hand side of eq. (\ref{eq:lhz}) is
\begin{align}
    N_c=\frac{1}{2}(N_l-1)(N_l-2).
\end{align}
The four-body term in eq. (\ref{eq:lhz}) consists of four neighboring qubits (three at the bottom boundary) as depicted in Fig. \ref{fig1}, and is therefore local and is possibly amenable to direct experimental implementation.

A recent contribution by Hartmann and Lechner \cite{hartmann2019rapid} showed that an infinite-range (mean-field) version of the four-body term  serves as a good approximation to the original nearest-neighbor (short-range) interactions, which greatly facilitates analytical studies. We therefore follow their idea and introduce the following Hamiltonian,
\begin{align}
\label{eq:p-spin}
\hat{H}_{P_2} = -\sum_{i=1}^{N} J_i \hat{\sigma}_i^z - N\left(\frac{1}{N} \sum_{i=1}^{N} \hat{\sigma}_i^z\right)^4.
\end{align}
This is the problem Hamiltonian we study in the present paper.

\section{Mean-field Analysis}
\label{section:MF_numerical}

It turns out to be convenient to introduce an additional parameter $\tau$ to control the time dependence of the constraint term.  The mean-field Hamiltonian is then written as
\begin{align}
\label{eq:p-spin_prime}
\hat{H}_{P_2^{\prime}}(s,\tau) &= -s \sum_{i=1}^{N} J_i \hat{\sigma}_i^z -\tau N\left(\frac{1}{N} \sum_{i=1}^{N} \hat{\sigma}_i^z\right)^4.
\end{align}
The total Hamiltonian is 
\begin{align}
\label{eq:hamiltonian}
\hat{H}(s,\tau) =\hat{H}_{P_2^{\prime}}(s,\tau) + (1-s) \hat{V}.
\end{align}
The parameters $s(t)$ and $\tau(t)$ are no longer linear in general as functions of $t$ and change from $s=\tau=0$ at $t=0$ to $s=\tau=1$ at $t=T$.

It is straightforward to apply the standard procedure to derive the free energy per qubit as a function of the ferromagnetic order parameter $m$ \cite{jorg2010energy,seki2012quantum,seoane2012,seki2015,hartmann2019rapid,susa2018quantum,ohkuwa2018reverse}.  We therefore just write the result for the free energy and its minimization condition, i.e. the self-consistent equation,
\begin{subequations}
\begin{align}
\label{eq:free_energy_finite_temp}
f(m) =& 3 \tau m^4 -\frac{1}{\beta} \left[ \ln 2\cosh \beta \sqrt{(4 \tau m^3+sJ_i)^2+(1-s)^2}\right]_i, \\
\label{eq:magnetization_finite_temp}
m =&\Biggl[ \frac{4 \tau m^3+sJ_i}{\sqrt{(4 \tau m^3+sJ_i)^2+(1-s)^2}}  \notag \\ 
&\times \tanh \beta \sqrt{( 4\tau m^3+sJ_i)^2+(1-s)^2}\Biggl]_i,
\end{align} 
\end{subequations}
where $\beta$ is the inverse temperature and the brackets $[\cdots ]_i$ stand for the average over the values of $J_i$, $\frac{1}{N}\sum_{i}\cdots$.

\begin{figure}[tb]
\centering
\subfigure[]{  
  \includegraphics[height = 5cm]{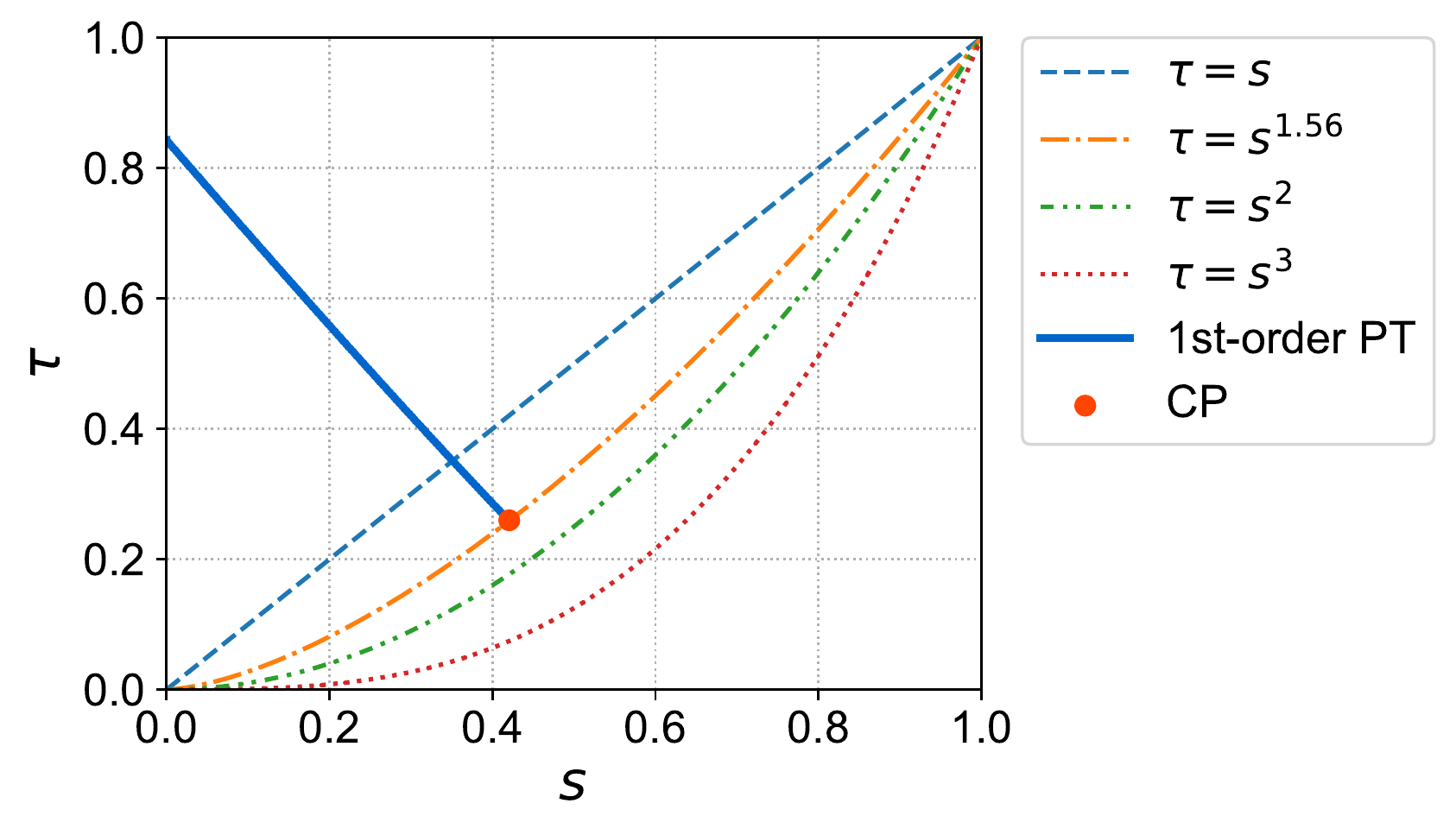}
  \label{fig2a}
  }
\subfigure[]{
  \includegraphics[height = 5cm]{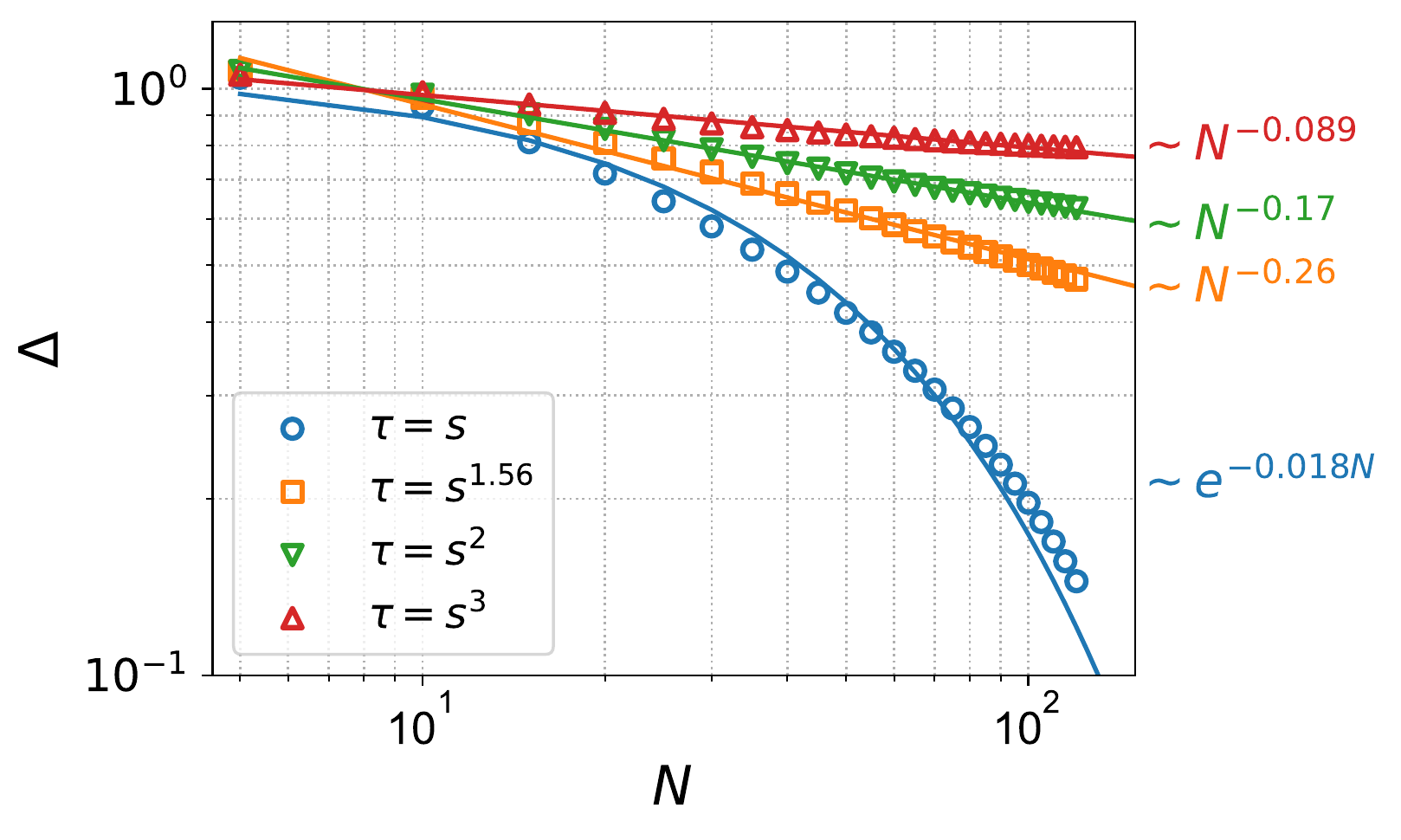}
  \label{fig2b}
}
\caption{(Color online) (a) Phase diagram of the Hamiltonian of eq. (\ref{eq:hamiltonian}) with eq. (\ref{eq:p-spin_prime}). The solid blue line denotes a line of first-order phase transitions (PT) and the orange dot represents the critical point (CP) of eqs. (\ref{eq:s_c}) and (\ref{eq:tau_c}). Each curve corresponds to the annealing  $\tau=s^r$ with four values of $r$.
(b) The minimum energy gap as a function of $N$ in a log-log scale. Full curves are fits to exponential or polynomial dependence. 
} 
\end{figure}

Let us first focus on the simplest case of zero temperature $\beta\to\infty$ and a uniform interactions $J_i=J$.  Then eqs.(\ref{eq:free_energy_finite_temp}) and (\ref{eq:magnetization_finite_temp}) reduce to
\begin{subequations}
\begin{align}
f(m) &= 3 \tau m^4- \sqrt{(4 \tau m^3+sJ)^2+(1-s)^2}, \\
\label{eq:magnetization}
m &=\frac{ 4\tau m^3+sJ}{\sqrt{(4 \tau m^3+sJ)^2+(1-s)^2}}.
\end{align}
\end{subequations}
Numerical solutions to these equations reveal the phase diagram on the $s$-$\tau$ plane for $J=0.5$ as shown in Fig. \ref{fig2a}, where the thick blue line denotes a line of first-order phase transitions terminating at a critical point marked in orange. The precise location of this critical point can be derived following the standard prescription that the derivatives up to third order should vanish at a critical point \cite{nishimori2010elements},
\begin{subequations}
\begin{align}
\label{eq:s_c}
s_c&=\frac{2^{5/2}}{3^{5/2} J +2^{5/2}}, \\
\label{eq:tau_c}
\tau_c &= \frac{J}{\sqrt{2}}\left(\frac{3^{5/2}-2^{5/2}}{3^{5/2} J +2^{5/2}}\right).
\end{align}
\end{subequations}

The conventional protocol of quantum annealing corresponds to the straight line $\tau=s$ in the phase diagram, which crosses the line of first-order phase transitions.  If we instead choose a trajectory $\tau =s^r$ with $r>1,56$, the annealing process does not encounter a phase transition.  The critical point is touched when $r=1.56$. Correspondingly, the minimum energy gap between the ground state and the first excited state closes exponentially as a function of the system size for $\tau=s$ whereas it is polynomial for $r=1.56$ as depicted in Fig. \ref{fig2b}. When $r>1.56$, the gap is expected to reach a constant in the large-$N$ limit because there is no phase transition, but the numerical data show a slow decay. This would probably due to the proximity of the curves $\tau=s^2$ and $\tau=s^3$ to the critical point and the asymptotic region for $N\gg 1$ is not yet reached.  It is anyway the case that an exponential speedup of the computation time can be achieved by the choice of $r\ge 1.56$ in comparison with the conventional annealing with $r=1$ because an exponential gap closing is avoided.

\begin{figure}[tb]
\centering
\subfigure[]{
  \includegraphics[height = 5cm]{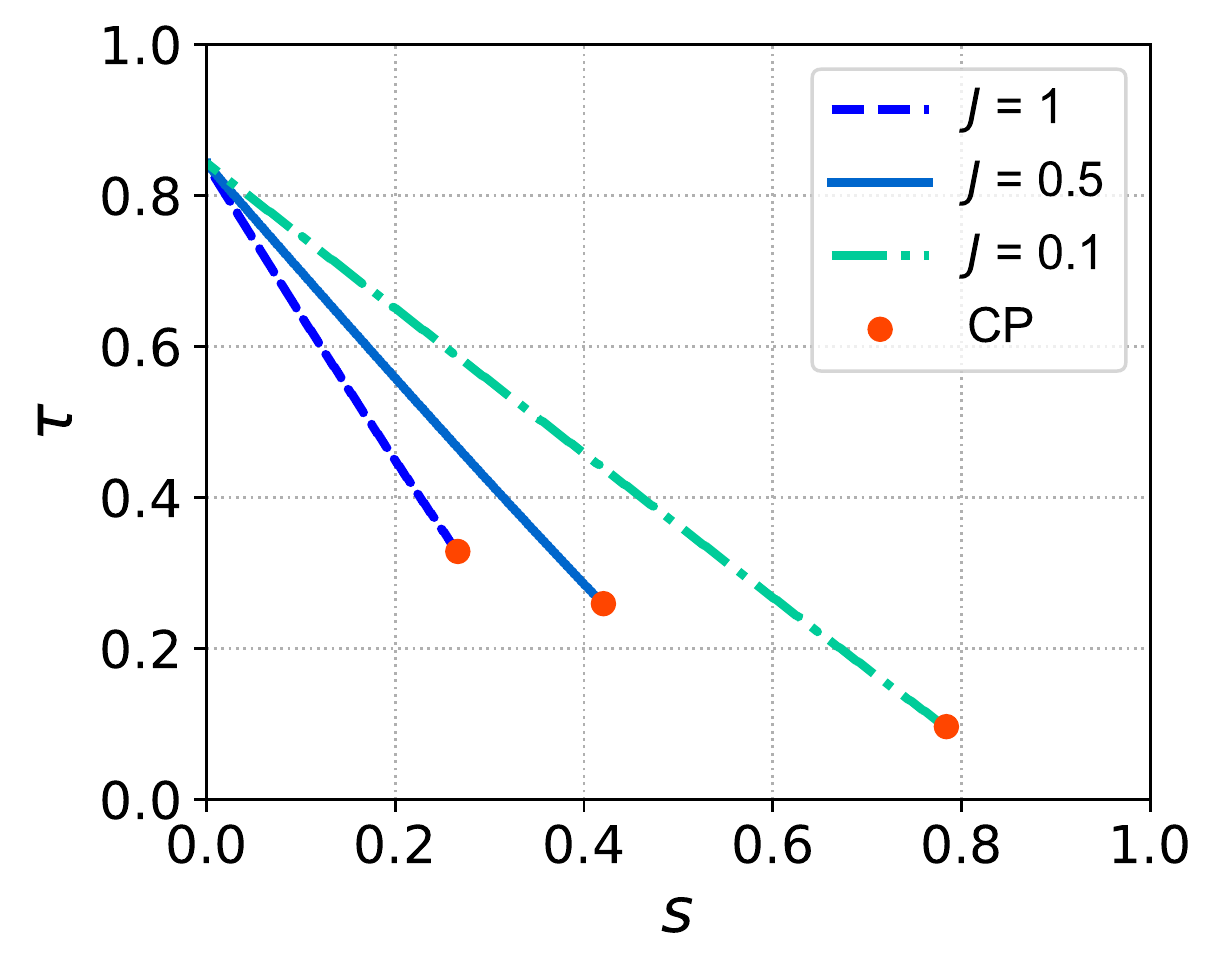}
  \label{fig3a}
}
\subfigure[]{
  \includegraphics[height = 5cm]{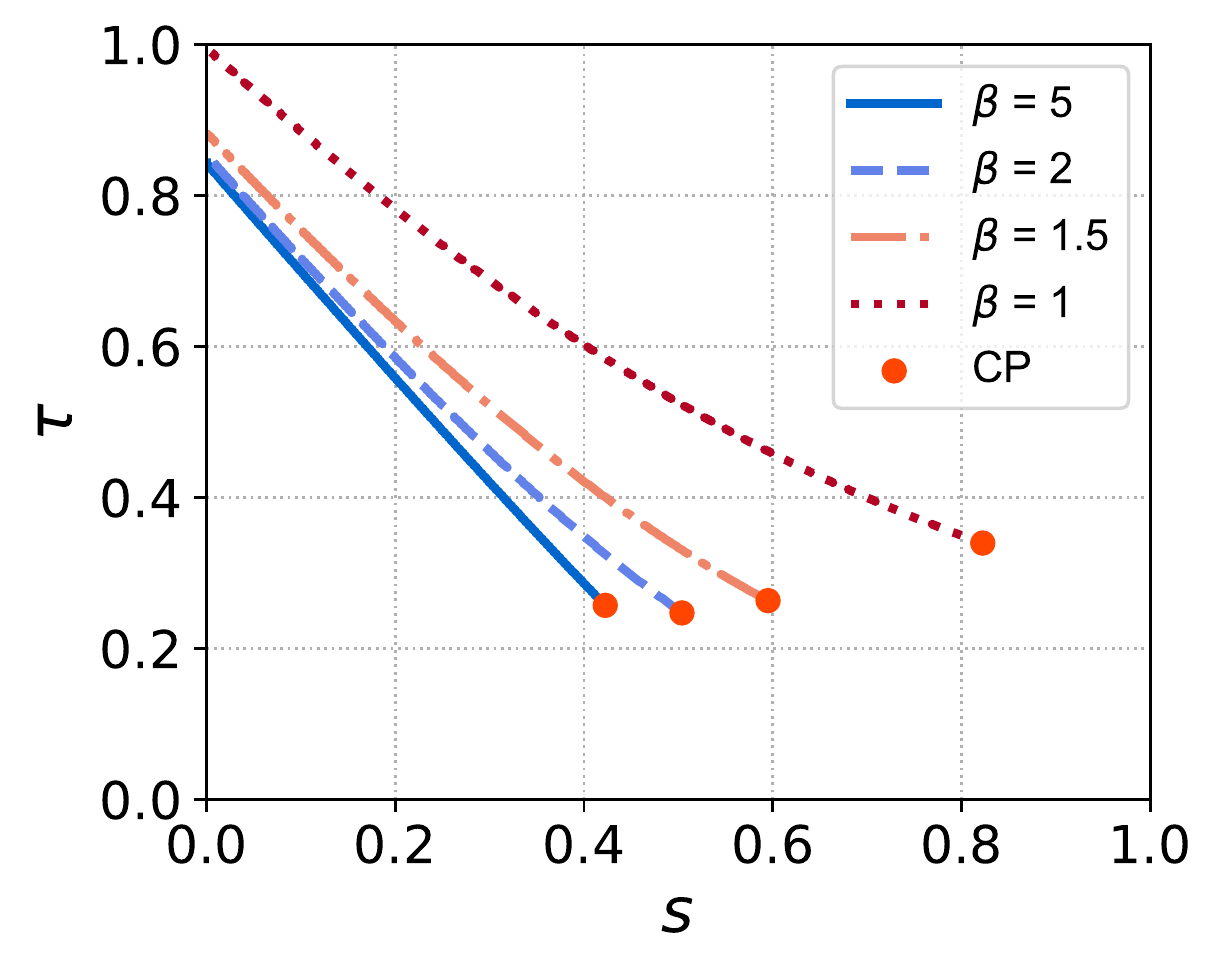}
  \label{fig3b}
}
\caption{
(Color online) (a) Phase diagram for $J=1,\ 0.5$, and $0.1$. (b) Phase diagram at finite temperatures. All lines are for first-order phase transitions and the orange dots indicate the critical point (CP).
}
\end{figure}

Similar results are obtained for different parameter values.  Examples are shown in Fig. \ref{fig3a} for $J=1, 0.5$, and 0.1, and Fig. \ref{fig3b} for $\beta=5, 2, 1.5$ and 1 with $J=0.5$.

\begin{figure}[tb]
\centering
\subfigure[]{  
  \includegraphics[height = 5cm]{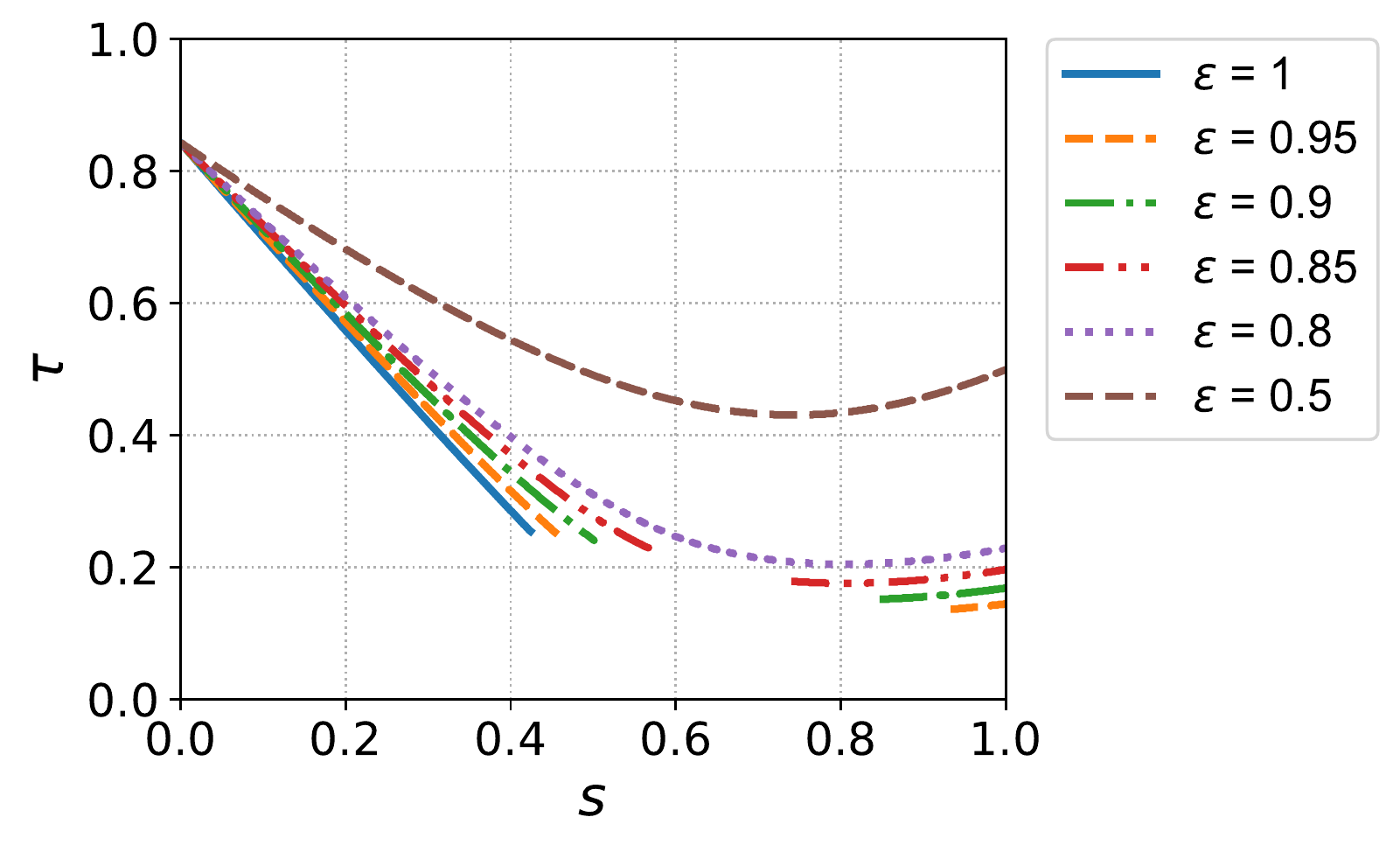}
  \label{fig4a}
  }
\subfigure[]{
  \includegraphics[height = 5cm]{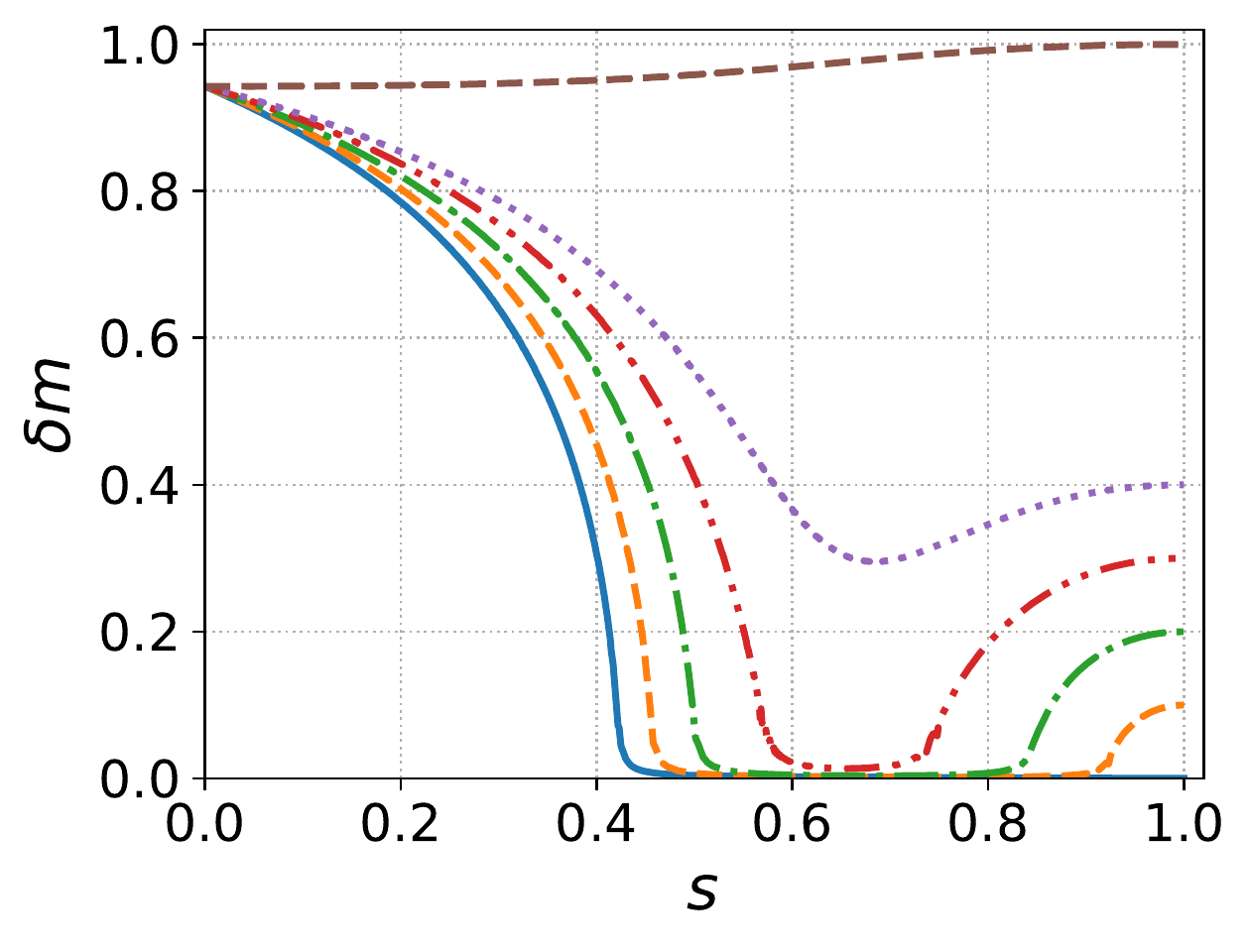}
  \label{fig4b}
}
\caption{(Color online) (a) Phase diagram for the mean-field model with randomness in the original interactions for the parameters $J=0.5$. Each curve indicates the line of  first-order phase transitions. (b) Jump in magnetization $m$ along the first-order transition line. The same color code is used for $\epsilon$ as in (a).}
\end{figure}

A more complex case of random interactions with the distribution function
\begin{align}
P(J_i)=\epsilon \delta(J_i-J)+ (1-\epsilon) \delta(J_i+J)\ \ \ (0\leq \epsilon \leq 1)
\end{align}
is interesting because this is for random and frustrated all-to-all interactions in the original problem, corresponding to the Sherrington-Kirkpatrick model of spin glasses \cite{Sherrington1975}. The ground-state free energy for this problem can be derived from eq. (\ref{eq:free_energy_finite_temp}) with the result
\begin{align}
f =& 3 \tau m^4- \epsilon \sqrt{\left( 4\tau m^3+s J\right)^2+\left(1-s\right)^2} \notag \\
&-(1-\epsilon)\sqrt{\left( 4\tau m^3-s J\right)^2+\left(1-s\right)^2}.
\end{align}
The phase diagram is drawn in Fig. \ref{fig4a} for a set of values of $\epsilon$.
It is observed that the lines of first-order phase transitions have breaks in the intermediate ranges of $s$ if $\epsilon$ is not close to 0.5. The latter is reasonable because $\epsilon=0.5$ represents a completely random spin-glass model, which is known as a very difficult problem to solve \cite{nishimoriSGbook}.  Nevertheless, even when the line of first-order transitions traverses the phase diagram as in the case of $\epsilon=0.8$, the jump in magnetization across a first-order transition can be tuned much smaller than the naive case of $\tau=s$ by an ingenious choice of the trajectory in the phase diagram as seen in Fig. \ref{fig4b}, which shows the magnetization jump along the first-order transition line.  This implies that quantum tunneling probability, which strongly depends on the width of an energy barrier represented by the magnetization jump, can be tuned to be larger by an appropriate choice of a trajectory connecting $s=\tau=0$ and $s=\tau=1$.  The extreme case of $\epsilon=0.5$ has no such properties.

\section{Conclusion}
\label{section:conclusion}
We have studied the scheme of Lechner, Hauke and Zoller \cite{lechner2015quantum} to express long-range (all-to-all), two-body interactions by short-range, many-body interactions.  Mean-field method was used to show that non-linear driving of the four-body constraint term as a function of time is advantageous for improved performance.  In particular, increasing the amplitude of the constraint term more slowly than for the intrinsic problem term can lead to an exponential speedup according to the mean-field prediction for the phase diagram although it would be difficult in practice to observe such a drastic effect because of non-ideal environmental effects as well as due to the limited applicability of mean-field theory. 

Since $\tau=s^r$ with $r>1.56$ means that the coefficient $\tau$ of the four-body constraint term increases slowly than that of the main problem term $s$, we may learn a generic lesson that constraints are better introduced later in the process of quantum annealing compared to the main problem Hamiltonian. In other words, one may first search for good solutions without constraints and then gradually select among candidate solutions those that satisfy the constraint. It is an interesting future topic to test this idea for various problems.

\section*{Acknowledgments}
This paper is based on results obtained from a project commissioned by the New Energy and Industrial Technology Development Organization (NEDO).  We have used QuTiP python library~\cite{johansson2012qutip,johansson2013qutip} in some of the calculations.

\end{document}